\documentclass[aps,prl,longbibliography,twocolumn,superscriptaddress,amsfont,graphicx,nofootinbib,preprintnumbers]{revtex4-1}%
\UseRawInputEncoding
\usepackage{color,graphicx,epsfig}
\usepackage{ifpdf}
\usepackage{amsmath}
\usepackage{bm}
\usepackage[english]{babel}
\usepackage{amssymb}
\usepackage{braket}
\usepackage{setspace}
\allowdisplaybreaks[4]

\usepackage{hyperref}
\usepackage{enumerate}
\usepackage{lipsum}

\usepackage{svg}
\svgsetup{
    inkscapepath=i/svg-inkscape/
}
\svgpath{{svg/}}
\usepackage{xcolor}
\usepackage{setspace}
\usepackage{hyperref}

\usepackage[utf8]{inputenc}
\usepackage[T1]{fontenc}
\usepackage{lmodern}

\usepackage{booktabs}
\usepackage{color}
\usepackage{epsfig}
\usepackage{ifpdf}
\usepackage{amsmath}
\usepackage{bm}
\usepackage[english]{babel}
\usepackage{amsfonts}
\usepackage{amssymb}
\usepackage{braket}
\usepackage{enumerate}
\usepackage{cancel}
\usepackage{multirow}
\usepackage{cleveref}
\usepackage{xspace}
\usepackage{array}
\usepackage{fancyvrb}
\usepackage{fontawesome}
\usepackage{dcolumn}
\usepackage{slashed}
\usepackage[normalem]{ulem}
\usepackage{url}
\usepackage{lineno}

\def\XXint#1#2#3{{\setbox0=\hbox{$#1{#2#3}{\int}$}
     \vcenter{\hbox{$#2#3$}}\kern-.5\wd0}}

\makeatletter
\g@addto@macro\bfseries{\boldmath}
\makeatother

\definecolor{nicered}{rgb}{0.7,0.1,0.1}
\definecolor{nicegreen}{rgb}{0.1,0.5,0.1}
\hypersetup{colorlinks, urlcolor=blue, citecolor=nicegreen,linkcolor= nicered}

\usepackage[absolute,overlay]{textpos}

\begin{document}

\begin{textblock}{4}(14,0.1)   
\includegraphics[width=2cm]{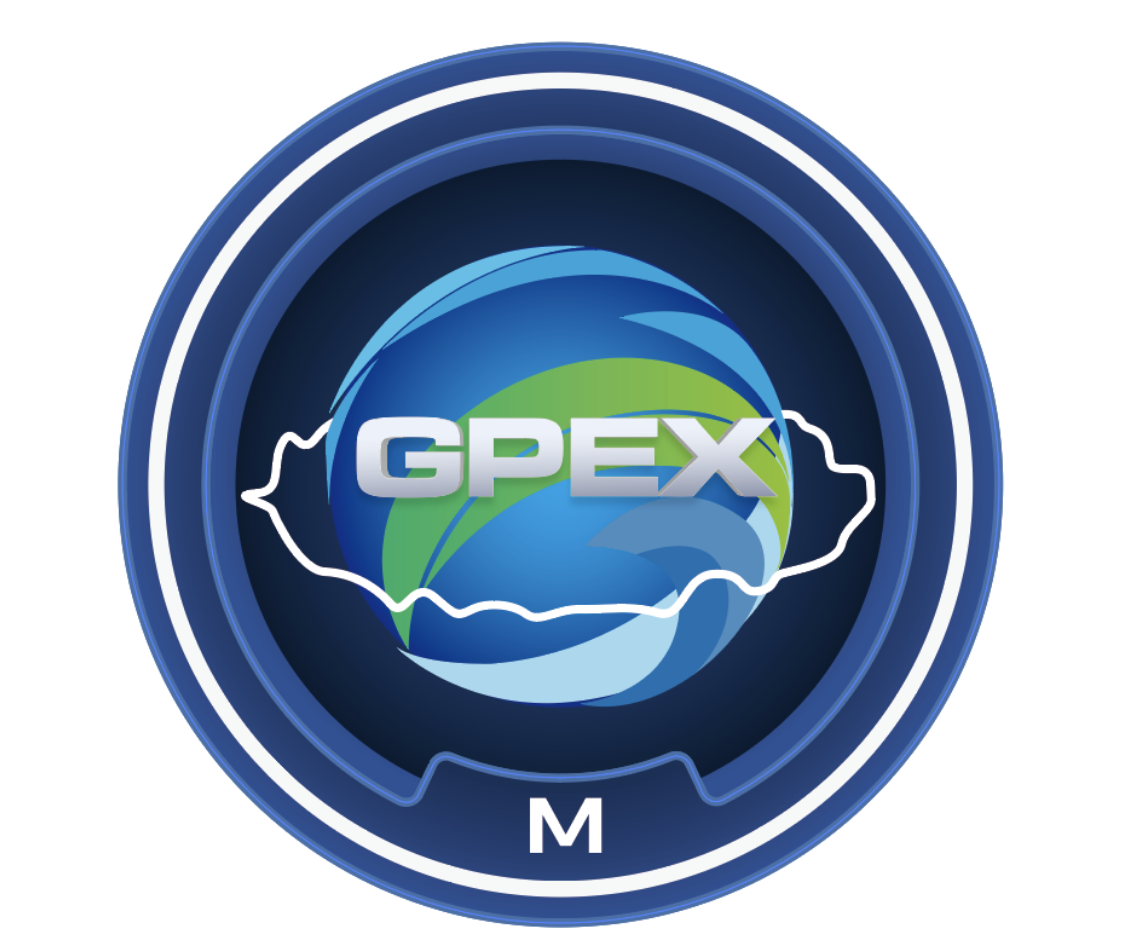}  
\end{textblock}


\title{Search for Ultralight Dark Matter with Quantum Magnetometry in the Earth’s Cavity} 

\author{Ariel Arza}
\altaffiliation{These authors contribute equally}
\affiliation{Department of Physics and Institute of Theoretical Physics, Nanjing Normal University, Nanjing, 210023, China}
\affiliation{Nanjing Key Laboratory of Particle Physics and Astrophysics, Nanjing, 210023, China}

\author{Yuanlin Gong}
\altaffiliation{These authors contribute equally}
\affiliation{Department of Physics and Institute of Theoretical Physics, Nanjing Normal University, Nanjing, 210023, China}

\author{Jun Guo}
\altaffiliation{These authors contribute equally}
\affiliation{College of Physics and Communication Electronics, Jiangxi Normal University, Nanchang 330022, China}

\author{Xiaofei Huang}
\altaffiliation{These authors contribute equally}
\affiliation{School of Instrumentation and Optoelectronic Engineering, Beihang University, Beijing 100191, China}

\author{Jing Shu}
\email{corresponding author: jshu@pku.edu.cn}
\affiliation{School of Physics and State Key Laboratory of Nuclear Physics and Technology, Peking University, Beijing 100871, China}
\affiliation{Center for High Energy Physics, Peking University, Beijing 100871, China}
\affiliation{Beijing Laser Acceleration Innovation Center, Huairou, Beijing, 101400, China}

\author{Hongliang Tian}
\affiliation{Department of Physics and Institute of Theoretical Physics, Nanjing Normal University, Nanjing, 210023, China}

\author{Wenyu Wang}
\affiliation{Beijing University of Technology, 100124, Beijing, China}

\author{Kai Wei}
\email{corresponding author: weikai@buaa.edu.cn}
\affiliation{Quantum Science and Technology College, Beihang University, Beijing 100191, China}
\affiliation{School of Instrumentation and Optoelectronic Engineering, Beihang University, Beijing 100191, China}

\author{Lei Wu}
\email{corresponding author: leiwu@njnu.edu.cn (project CI)}
\affiliation{Department of Physics and Institute of Theoretical Physics, Nanjing Normal University, Nanjing, 210023, China}
\affiliation{Nanjing Key Laboratory of Particle Physics and Astrophysics, Nanjing, 210023, China}

\author{Mingming Xia}
\affiliation{Hangzhou Institute of Extremely-Weak Magnetic Field Major National Science and Technology Infrastructure, Hangzhou, 310051, China}

\author{Jin Min Yang}
\affiliation{Center for Theoretical Physics, Henan Normal University, Xinxiang 453007, China}
\affiliation{Institute of Theoretical Physics, Chinese Academy of Sciences, Beijing 100190, China}

\author{Qiang Yuan}
\affiliation{Key Laboratory of Dark Matter and Space Astronomy, Purple Mountain Observatory,
Chinese Academy of Sciences, Nanjing 210008, China}
\affiliation{School of Astronomy and Space Science, University of Science and Technology of China,
Hefei 230026, China}

\author{Yang Zhang}
\altaffiliation{These authors contribute equally}
\affiliation{Center for Theoretical Physics, Henan Normal University, Xinxiang 453007, China}

\author{Yi Zhang}
\affiliation{Key Laboratory of Dark Matter and Space Astronomy, Purple Mountain Observatory,
Chinese Academy of Sciences, Nanjing 210008, China}
\affiliation{School of Astronomy and Space Science, University of Science and Technology of China,
Hefei 230026, China}

\author{Bin Zhu}
\email{corresponding author: zhubin@mail.nankai.edu.cn}
\affiliation{School of Physics, Yantai University, Yantai 264005, China}

\begin{abstract}
Ultralight dark matter candidates, such as axions and dark photons, are leading dark matter candidates. They may couple feebly to photons, sourcing oscillating electromagnetic signals in the Earth's conducting cavity formed between the ground and the ionosphere, providing detectable magnetic field signatures at wavelengths above the Earth's size. We carry out a project aiming to search for new physics using an unshielded high-sensitivity atomic magnetometer, termed the Geomagnetic Probe for nEw physiCS (GPEX). In this work, we report our first search for axion and dark photon dark matter, conducted in the desert of XiaoDushan in Gansu Province, China. Analysis of the collection of one-hour data shows no robust evidence for axion- or dark photon-induced magnetic signals. Correspondingly, we set the constraints on the axion-photon coupling with $g_{a\gamma\gamma} < 7\times10^{-10}\, \mathrm{GeV^{-1}}$ and the dark photon kinetic-mixing parameter $\epsilon < 2\times10^{-6}$ in the mass range $3.5 \times 10^{-16}\, \mathrm{eV} \sim 1.8 \times 10^{-14}\, \mathrm{eV}$. Our findings demonstrate the feasibility of using ground-based quantum magnetic sensors for ultralight dark matter searches. Future networks of such detectors operating over extended periods could improve the sensitivity by about three orders of magnitude.


\end{abstract}


\maketitle

{\it \textbf {Introduction}.} Cosmological observations indicate that only about 16\% of the universe's total matter is baryonic. While the Standard Model of particle physics accurately describes this component, the nature of the remaining $\sim$84\%—the enigmatic dark matter—remains a fundamental problem at the intersection of particle physics, cosmology, and astrophysics \cite{Bertone:2010zza}. Consequently, ultralight dark matter (ULDM) candidates now arise from extensions to the Standard Model, such as axions, axion-like particles, and dark photons \cite{Preskill:1982cy,Abbott:1982af,Dine:1982ah,Nelson:2011sf,Arias:2012az,Jaeckel:2012mjv}. The axion, originally proposed to solve the strong-CP problem in quantum chromodynamics \cite{Peccei:1977hh}, also emerges naturally in high-energy frameworks such as string theory and extra-dimensional models \cite{Arvanitaki:2009fg,Svrcek:2006yi, Abel:2008ai}. Another promising candidate, the dark photon, provides a portal to the dark sector through kinetic mixing with the Standard Model photon \cite{Holdom:1985ag, Dienes:1996zr, Abel:2006qt, Goodsell:2009xc, Nelson:2011sf}. A key cosmological virtue of these ultralight fields is that they can be produced in the early universe via non-thermal mechanisms, naturally accounting for the observed dark matter relic density \cite{Preskill:1982cy, Abbott:1982af, Dine:1982ah, Nelson:2011sf, Arias:2012az, Graham:2015rva}.


ULDM with masses ranging from $10^{-22}\,\mathrm{eV}$ to a few $\mathrm{eV}$, behaves as a weakly coupled, coherent classical field that permeates all space. Detection of ULDM relies on their non-gravitational couplings to Standard Model particles \cite{Bradley:2003kg,Budker:2013hfa,An:2013yua,CAST:2017uph, Irastorza:2018dyq,ADMX:2019uok,Gramolin:2020ict,Smorra:2019qfx, Sikivie:2020zpn,Jiang:2021dby,Chen:2021bdr,XENON:2022ltv,Bloch:2022kjm,Wei:2023rzs,Jiang:2023jhl,Arza:2023rcs,An:2023wij, Kimball:2023vxk,Xu:2023vfn,Gavilan-Martin:2024nlo}. For instance, a suite of specialized experimental approaches has been developed to probe their couplings to the photon field. These include shielded experiments such as ADMX \cite{ADMX:2019uok}, DM-Radio \cite{DMRadio:2022pkf}, ABRACADABRA \cite{Salemi:2021gck}, and others involving  quantum sensing and their networks, such as atomic spectroscopy \cite{Berger:2022tsn}, superconducting cavities \cite{Dixit:2020ymh,Agrawal:2023umy,Cervantes:2022gtv,SHANHE:2023kxz,SHANHE:2024tpr,Zheng:2025qgv}, atomic magnetometers \cite{Afach:2021pfd,Jiang:2023jhl}, state-swapping techniques \cite{Chen:2021bgy,Wurtz:2021cnm,Chen:2023ryba,Jiang:2022vpm}, and squeezed-state receivers \cite{HAYSTAC:2020kwv}. Alternatively, the Earth's interior and ionosphere forms a natural cavity that functions as a giant transducer for ULDM. In this cavity, the conversion of dark matter into a detectable magnetic field is highly effective for signal frequencies below 7 Hz, which corresponds to dark matter masses less than $3\times10^{-14}\,\mathrm{eV}$ (Compton wavelengths larger than the Earth Radius). Competitive constraints on axion and dark photon dark matter have been derived from the analysis of long-term geomagnetic datasets from the SuperMAG collaboration \cite{Arza:2021ekq,Fedderke:2021aqo,Fedderke:2021rrm,Friel:2024shg} and the Eskdalemuir Observatory (UK) \cite{Nishizawa:2025xka,Nomura:2025rfi}. However, conventional long-term geomagnetic measurements usually suffer from systematics and are not optimized for dark matter searches. On the other hand, the SNIPE-Hunt collaboration has further advanced the search by setting constraints with a dedicated network of classical magnetometers \cite{Sulai:2023zqw,highlight}. Building on this idea, future efforts can extend to different search strategies \cite{Bloch:2023wfz} and the probing of other 
ULDM candidates such as millicharged particles \cite{Arza:2025cou}.

\begin{figure}[ht]
    \centering
    \includegraphics[width=1\linewidth]{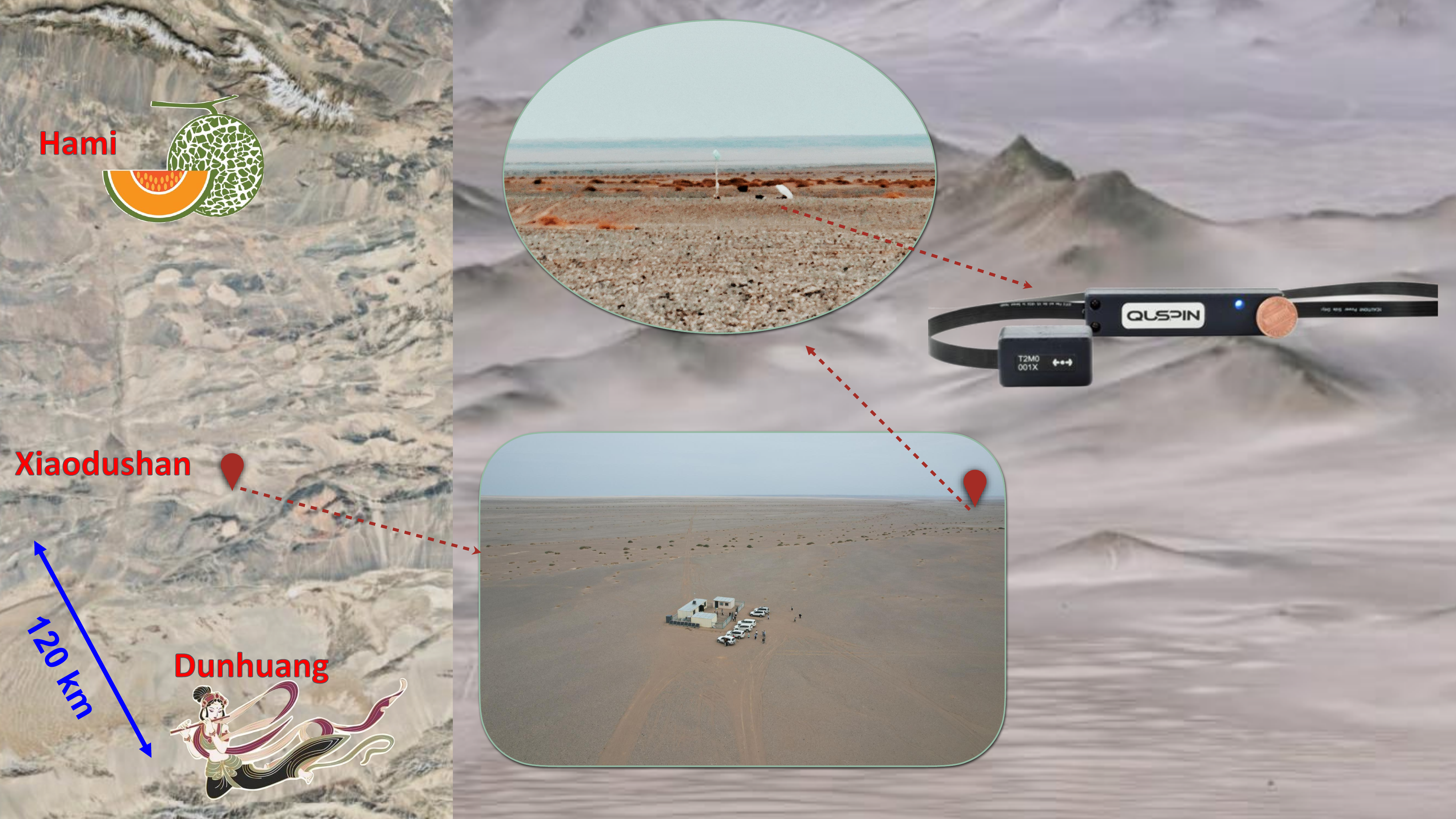}
    \caption{Setup for our experiment. Away from the Dunhuang city 120 km, in the desert of XiaoDushan, we made our one-hour high performance measurement using the QTFM-B scalar atmoic magnetometer.}
    \label{fig:setup}
\end{figure}

This work reports the inaugural results of the Geomagnetic Probe for nEw physiCS (GPEX) experiment, which searches for geomagnetic signatures from ULDM candidates. For this first-stage campaign, we operated an unshielded, optically pumped rubidium vapor magnetometer in the electromagnetically pristine environment of the XiaoDushan desert in China. The choice of quantum sensor enables precise measurements and can be readily integrated into future networks of synchronized sensors, while the remote desert environment minimizes anthropogenic electromagnetic disturbances, enhancing our ability to detect subtle signals. Our analysis of one hour of data revealed no robust signals of axions or dark photons. We therefore place upper limits on their couplings to photons, achieving constraints of $g_{a\gamma\gamma} \sim 7\times10^{-10}\, \mathrm{GeV^{-1}}$ for axions and $\epsilon \sim 2\times10^{-6}$ for dark photons, which surpass those from the previous SNIPE-Hunt collaboration by an order of magnitude and a factor of three, respectively. This demonstrates the powerful potential of unshielded atomic magnetometers for this search.

{\it \textbf {Theoretical background}.} Ultralight dark matter can interact feebly with electromagnetic fields. Because its occupancy number is extremely high, the dark matter field behaves classically. This collective effect can be incorporated into Maxwell's equations as an effective classical current density $\vec J_\mathrm{DM}$, which acts as a source for electromagnetic phenomena. In this context, the phenomenology is reduced to solve the classical equations \cite{Sikivie:2020zpn}
\begin{align}
\vec\nabla\cdot\vec E & = 0 \label{eq:Gauss}
\\
\vec\nabla\cdot\vec B & = 0 \label{eq:mon}
\\
\vec\nabla\times\vec E & = -\partial_t\vec B \label{eq:Far}
\\
\vec\nabla\times\vec B & = \partial_t\vec E+\vec J_\mathrm{DM} ~, \label{eq:Amp}
\end{align}
where $\vec B$ and $\vec E$ are the magnetic and electric fields, respectively. Effective dark matter charge densities are suppressed by the dark matter velocity.

We use spherical coordinates $\vec x=(r,\theta,\varphi)$ and expand the effective current $\vec J_\mathrm{DM}$ in series of vector spherical harmonics (VSH), $\vec Y_{\ell m}$, $\vec\Psi_{\ell m}$ and $\vec\Phi_{\ell m}$ as
\begin{align}
\vec J_\mathrm{DM}(\vec x,t) &= e^{-i\omega t}\sum_{\ell,m}\left(J_{\ell m}^{(r)}(r)\vec Y_{\ell m}(\theta,\varphi)+J_{\ell m}^{(1)}(r)\vec\Psi_{\ell m}(\theta,\varphi)\right. \nonumber
\\
& ~~~~~~~~~~~~~~~~~ \left.+J_{\ell m}^{(2)}(r)\vec\Phi_{\ell m}(\theta,\varphi)\right) ~, \label{eq:VSHexp}
\end{align}
where $\ell=0, 1, 2,...$ and $m=-\ell,-\ell+1,...,\ell-1,\ell$. The parameter $\omega$ corresponds to the typical angular frequency of the system, which is equivalent to the dark matter particle masses $m_a$ and $m_{\gamma'}$ for axions and dark photons, respectively. For ultralight fields such as axions and dark photons, the identity $\vec\nabla\times\vec J_\mathrm{DM}=0$ is satisfied. From VSH properties, it is implied directly that $J_{\ell m}^{(2)}=0$, and that $J_{\ell m}^{(r)}=r {J_{\ell m}^{(1)\prime}}+J_{\ell m}^{(1)}$. We look for magnetic field solutions at the Earth's surface, considering that electromagnetic fields are bounded on the one side by the ground and on the other side by the ionosphere. For Compton wavelengths bigger than the ionosphere height, the dark matter induced magnetic field signal $\vec B_\mathrm{DM}$ at the Earth's surface is determined uniquely by the radial component $J_{\ell m}^{(r)}$ of the effective current. We find
\begin{equation}
\vec B_\mathrm{DM}(\theta,\varphi,t) = -e^{-i\omega t}\sum_{\ell,m}{R_e J_{\ell m}^{(r)}(R_e)\vec\Phi_{\ell m}(\theta,\varphi)\over\ell(\ell+1)-\omega^2R_e^2} ~, \label{eq:gensignal}
\end{equation}
where $R_e=6371.2\,\mathrm{km}$ is the Earth's radius. The divergences shown in Eq. (\ref{eq:gensignal}) correspond to Schumann resonances. These resonances are not discussed in this work since the target parameter space satisfies $\omega^2R_e^2<2$.

For axions, the effective current at $r=R_e$ is \cite{Arza:2021ekq,Taruya:2025bhe}
\begin{equation}
\vec J_a=ig_{a\gamma\gamma}m_aa_0e^{-im_at}\sum_{\ell,m}C_{\ell m}\left((\ell+1)\vec Y_{\ell m}-\vec\Psi_{\ell m}\right) ~, \label{eq:effcurraxion}
\end{equation}
where $g_{a\gamma\gamma}$ is the axion-photon coupling constant, $a_0$ the complex axion amplitude, normalized by ${1\over2}m_a^2\left<|a_0|^2\right>=\rho_\mathrm{0}$, where $\rho_\mathrm{0}$ is the local dark matter energy density which in this work is taken as $0.3\,\mathrm{GeV/cm}^3$. The coefficients $C_{\ell m}$ are taken from the IGRF-13 model of the Earth's geomagnetic field \cite{alken2021international}. By comparing Eq. (\ref{eq:effcurraxion}) with the general expression in Eq. \eqref{eq:VSHexp}, the radial component $J_{\ell m}^{(r)}$ of the effective current is identified, thus the axion signal is computed as
\begin{equation}
\vec B_a=-ig_{a\gamma\gamma}m_a R_e\,a_0\,e^{-im_at}\sum_{\ell,m}{(\ell+1)C_{\ell m}\vec\Phi_{\ell m}(\theta,\varphi)\over \ell(\ell+1)-m_a^2R_e^2} ~. \label{eq:axionsignal}
\end{equation}

For dark photons, the effective current is given by \cite{Fedderke:2021aqo}
\begin{equation}
\vec J_{\gamma'}=-\sqrt{4\pi\over3}\epsilon m_{\gamma'}^2\sum_{m=-1}^1 A_m'\left(\vec Y_{1m}+\vec\Psi_{1m}\right)e^{-i\omega_mt} ~, \label{eq:effcurrdp} 
\end{equation}
where $\epsilon$ is the kinetic mixing parameter between dark photons and standard model photons and $A_m'$ the dark photon amplitudes defined in terms of $\vec A'=(A_x',A_y',A_z')$ as $A_1=-{1\over\sqrt{2}}(A_x'-iA_y')$, $A_{-1}={1\over\sqrt{2}}(A_x'+iA_y')$ and $A_0'=A_z'$, where $\vec A'$ is normalized as ${1\over2}m_{\gamma'}^2\left<|\vec A|^2\right>=\rho_\mathrm{0}$. For dark photons, the oscillation frequency is defined as $\omega_m=m_{\gamma'}-2\pi m f_d$, where $f_d$ is the frequency associated to the sidereal day, which accounts for the effects of Earth's rotation. As the data used in this work correspond to a couple of hours, we neglect Earth's rotation effects. Just like in the axion case, we identify $J_{\ell m}^{(r)}$ in Eq. (\ref{eq:effcurrdp}) and find that the dark photon magnetic field signal is given by
\begin{equation}
\vec B_{\gamma'}= \sqrt{4\pi\over3}{\epsilon m_{\gamma'}^2R_e\over2-m_{\gamma'}^2R_e^2}\,e^{-im_{\gamma'}t}\sum_{m=-1}^1 A_m'\vec\Phi_{1 m}(\theta,\varphi) ~. \label{eq:dpsignal}   
\end{equation}

The dark matter induced magnetic field is an almost monochromatic signal with frequency $\nu=\omega/(2\pi)$. The bandwidth of this signal is determined by the velocity distribution of the dark matter particles. The dark matter velocity dispersion is taken conventionally as $\Delta v\sim10^{-3}$, leading to a bandwidth signal of $\Delta\nu=v\Delta v\nu\sim10^{-6}\,\nu$. At the same time, within a time scale smaller than the coherent time $\tau = 1/\Delta \nu \approx 278\,\mathrm{hour}\left( \frac{\mathrm{Hz}}{\nu} \right) $, the dark matter field amplitudes $a_0$ or $A_i$ (Eq. (\ref{eq:axionsignal}) and (\ref{eq:dpsignal})) become stochastically fluctuating \cite{Derevianko:2016vpm,Centers:2019dyn,Chen:2021bdr,Nakatsuka:2022gaf}. As shown later, since our measurement time is much shorter than the coherent time, we can ignore the bandwidth when incorporating this stochastic effect into the analysis \cite{Budker:2013hfa,OHare:2017yze,Centers:2019dyn}.

{\it \textbf{Experimental set-up and data analysis}.} In our experiment, we employed the QTFM-B scalar atomic magnetometer manufactured by QUSPIN \cite{magnetometer}. This instrument is a pulsed, optically pumped rubidium magnetometer that operates based on the Free Induction Decay (FID) detection scheme. The vapor cell contains rubidium atoms that are optically pumped to a polarized state. In the presence of an external magnetic field, these atoms undergo Larmor spin precession at a frequency directly proportional to the magnitude of the applied field. This precession alters the optical absorption and dispersion properties of the atomic vapor, which we detect by measuring the transmission of a 795 nm vertical-cavity surface-emitting laser through the atomic vapor during the FID signal. The QTFM-B magnetometer can be used to perform a scalar magnetic field measurements at frequencies below 500 Hz. It offers a magnetic field sensitivity below $20\,\mathrm{pT}/\sqrt{\mathrm{Hz}}$ and supports a maximum data acquisition rate of 1000 samples per second.

The experiment was conducted on 7 August 2025 in the desert of XiaoDushan, located approximately 120 km from Dunhuang City, Gansu Province, China (geographic coordinates: 40.9664° N, 93.9822° E, altitude 1268 m above sea level), as illustrated in Fig. \ref{fig:setup}. During the measurement, the ambient temperature was recorded at $27\,^{\circ}\mathrm{C}$, which is within the operational range of the magnetometer. Data collection took place between 12:30 pm and 3:30 pm local time, yielding an effective observation time of $T_{\rm{obs}} = 1.05\,\mathrm{h}$ at a sampling frequency of 62.5 Hz. The magnetometer was connected to a laptop via a 2 m USB cable and controlled using dedicated software that enabled configuration of the sampling rate and calibration of the sensor orientation to avoid the instrument’s dead zone. A trial measurement of approximately 30 minutes was first performed to verify the sensor alignment, sampling functionality, and data recording. Subsequently, the main data run was carried out, and the resulting time-series data were recorded on the laptop.



We analyze the data by searching for excess power in the frequency domain, as characterized by the power spectral density (PSD) \cite{1975ApJS...29..285G,Derevianko:2016vpm,Fedderke:2021rrm}. The remote desert location of the experiment minimizes dominant anthropogenic noise contributions at low frequencies. The primary source of noise in the recorded data is the magnetometer itself, which we assume to exhibit both Gaussian and stationary behavior throughout the measurement period. Below approximately 0.5 Hz, the noise follows a $1/f$ power-law dependence, while at higher frequencies, it becomes approximately frequency-independent. Accordingly, we restrict our analysis to the frequency range between 0.1 Hz and 5 Hz. This range is chosen to ensure reliable extraction of magnetic field signals under the condition $(\omega R_e)^2 < 2$, where $\omega$ is the angular frequency and $R_e$ is the Earth’s radius. Within this band, the noise variance is estimated from the power in adjacent frequency bins at each frequency. This yields a nearly flat noise background, providing an optimal regime in which narrow spectral features—such as potential signals from ultralight dark matter—can be identified and studied.


The amplitude and thus the PSD of a potential signal at a given frequency is fully determined by Eq. (\ref{eq:axionsignal}) and (\ref{eq:dpsignal}), and can be identified within a background of approximately flat noise. The intrinsic bandwidth of the signal is approximately $5 \times 10^{-6} \, \mathrm{Hz}$, which is much smaller than the width of a single frequency bin. Specifically, each frequency bin has a resolution of $1/T_{\mathrm{obs}} = 2.8 \times 10^{-4} \, \mathrm{Hz}$, ensuring that the entire signal bandwidth is contained within a single bin. Given the relatively short observation time, $T_{\mathrm{obs}} \ll \tau = 55.6 \, \mathrm{hours} \left( \frac{5 \, \mathrm{Hz}}{\nu} \right)$, we model the expected dark matter signal as a single narrow peak appearing in each relevant frequency bin \cite{Budker:2013hfa}.

\begin{figure}[ht]
    \centering
    \includegraphics[width=1\linewidth]{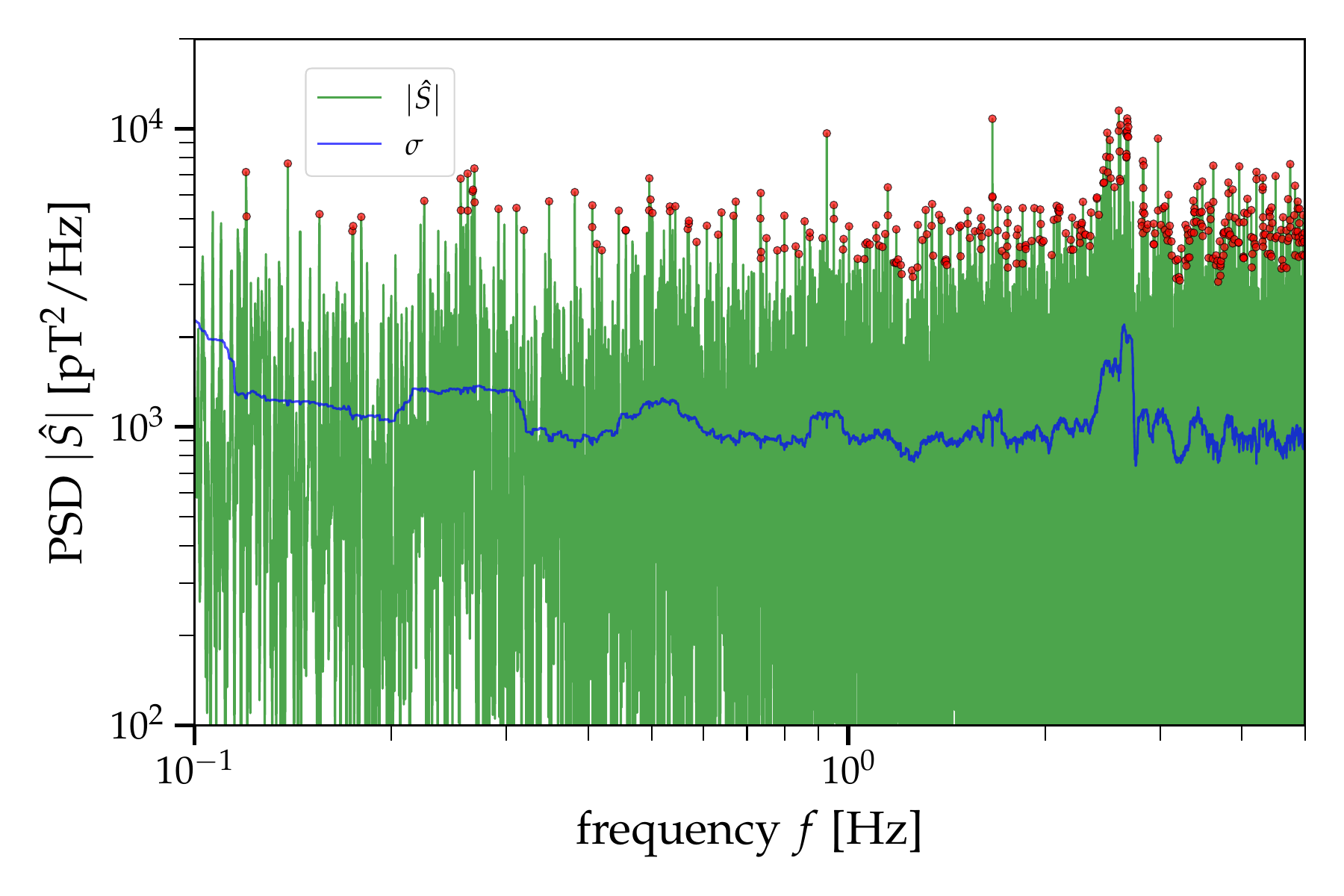}
    \caption{PSD of the observed magnetic fields, $\hat{S}$ in green. and the standard deviation, $\sigma$ in blue, evaluated from 200 bins in the immediate vicinity on both sides excluding three center bins. We also marked the naive signal candidates with ${\rm SNR}\geq 2$ by circles in red.}
    \label{fig:psd}
\end{figure}

\begin{figure}[ht]
    \centering
    \includegraphics[width=1\linewidth]{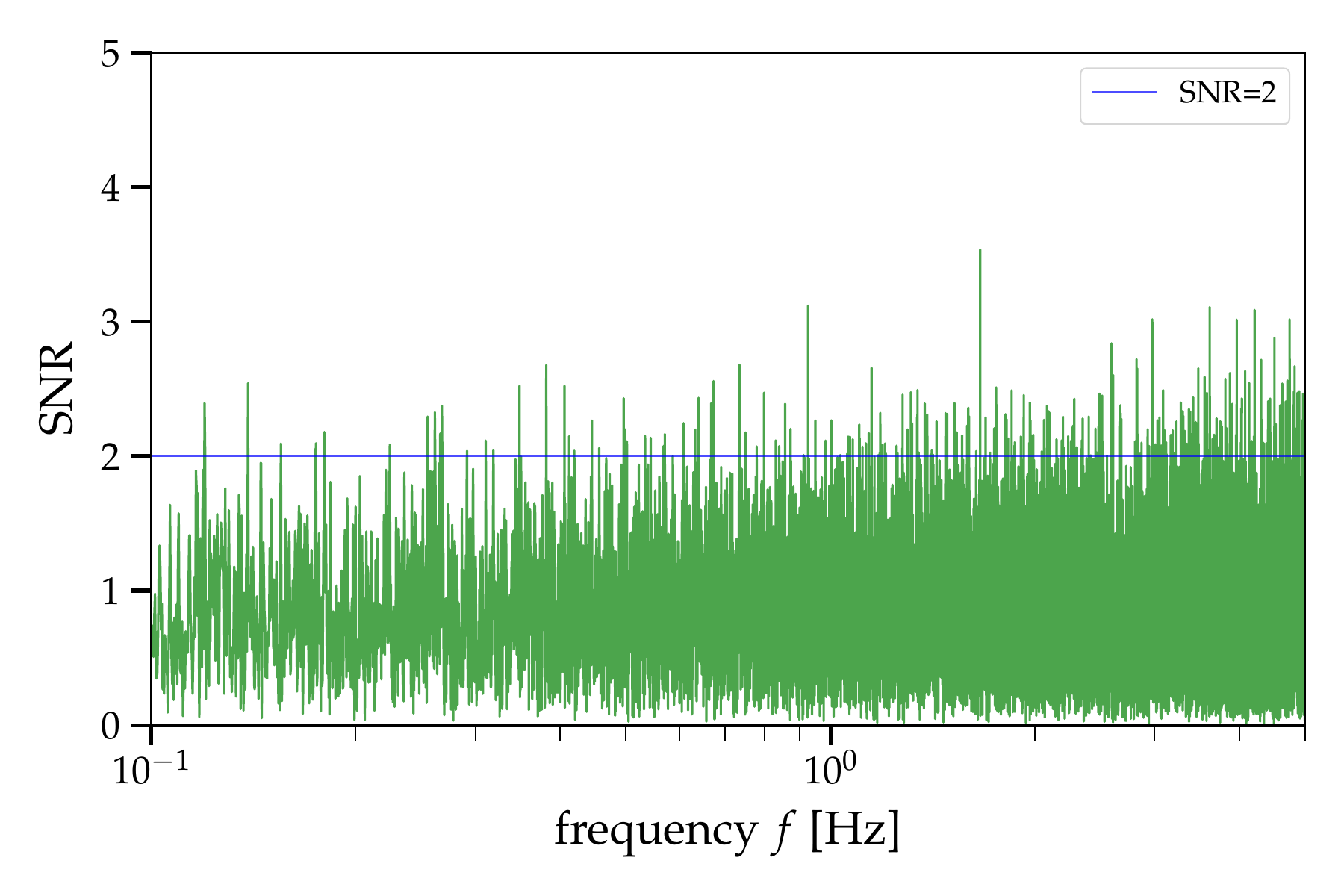}
    \caption{SNR for the frequency of interest. The horizontal solid lines is SNR= 2.}
    \label{fig:snr}
\end{figure}

We neglect any cross talk between the PSDs of the signal and noise across all frequencies, as their underlying sources are uncorrelated. Consequently, the single-sided PSD of the measured magnetic field $\vec{B}(t) = \vec{B}_{\rm n}(t) + \vec{B}_{\rm s}(t)$ at frequency $f$ is given by the sum of the noise and signal contributions \cite{Nishizawa:2025xka}:
\begin{align}
    S(f) &= S_{\rm n}(f) + S_{\rm s}(f) \nonumber \\ 
         &= S_{\rm n}(f) + \frac{2}{T_{\rm obs}} \left| \vec{R}(f_{\rm s}, \mathcal{A}_{\rm s}) \right|^2 g_{\rm s}^2 \left\{ \delta(f - f_{\rm s}) \right\}^2 \;,
    \label{eq:psd}
\end{align}
where $S_{\rm n}(f)$ is the PSD of the noise magnetic field $\vec{B}_{\rm n}(t)$ over the total observation time $T_{\rm obs}$, and $S_{\rm s}(f)$ is the PSD of the signal magnetic field $\vec{B}_{\rm s}(t)$, with the signal type denoted by $\mathrm{s} = a, \gamma'$ (corresponding to axion and dark photon, respectively). The response function $\vec{R}(f, \mathcal{A})$ relates the signal field to the time-domain representation
\begin{equation*}
    \vec{B}_{\rm s}(t) \equiv \vec{R}(f_{\rm s}, \mathcal{A}_{\rm s}) \, g_{\rm s} \, \exp\left(-2\pi i f_{\rm s} t \right) \;,
\end{equation*}
with $g_{a, \gamma'} = g_{a\gamma\gamma}, \epsilon$ and $\mathcal{A}_{a, \gamma'} = a_0, A'$ representing the coupling constants and field amplitudes for axions and dark photons, respectively. The delta function $\delta(f - f_{\rm s})$ accurately models the line-like spectral signature of the signal for our relatively short observation time $T_{\rm obs}$, which is much smaller than the coherent integration time $\tau$. Since the noise is statistically stationary within the measurement window in the frequency domain, we define an estimator for the dark matter line signal at frequency $f = f_{\rm s}$ as
\begin{equation}
    \hat{S}(f_{\rm s}) \equiv S_{\rm n}(f_{\rm s}) + 2 T_{\rm obs} \left| \vec{R}(f_{\rm s}, \mathcal{A}_{\rm s}) \right|^2 g_{\rm s}^2 \;,
    \label{eq:estimator}
\end{equation}
where we have approximated $\delta(0) \approx T_{\rm obs}$ for finite observation time. The noise PSD, $S_{\rm n}(f_{\rm s})$, follows a $\chi^2$ distribution with two degrees of freedom \cite{1975ApJS...29..285G}.  For cases where the observation time exceeds the coherence time, we refer the reader to Refs. \cite{Fedderke:2021rrm,Nishizawa:2025xka} for the definition of the weighted estimator applied over multiple data segments, which incorporates the signal bandwidth and corresponding corrections. Moreover, we provide representative examples of the noise amplitude distributions at selected frequencies in the Supplementary Material, demonstrating that they are well described by a $\chi^2$ distribution as expected. The $p$-values from the $\chi^2$ goodness-of-fit test across the entire frequency range of interest are also shown.


We first identify potential signal candidates by calculating the signal-to-noise ratio (SNR), defined as
\begin{equation}
    {\rm SNR} \equiv \sqrt{\frac{\langle \hat{S}(f_{\rm s}) \rangle}{\sigma(f_{\rm s})}} \;,
    \label{eq:snr_def}
\end{equation}
where $\sigma(f_{\rm s})$ is the standard deviation of the noise and $\langle \cdot \rangle$ denotes the ensemble average. The noise standard deviation $\sigma(f_{\rm s})$ is evaluated using the observed PSD over $\pm 200$ frequency bins centered around $f = f_{\rm s}$, excluding the three bins immediately surrounding $f = f_{\rm s}$ to avoid contamination from a potential signal. This approach ensures that the noise variance estimate is independent of any candidate signal itself and automatically excludes signals broader than a single frequency bin \cite{Nishizawa:2025xka}.

In Fig.~\ref{fig:psd}, we present the amplitude of $|\hat{S}(f)|$ along with its standard deviation across the frequency range. Across the mass ranges of interest, we identify approximately 300 naive candidates with ${\rm SNR} \geq 2$, but find no candidate exceeding ${\rm SNR} \geq 5$ - a threshold commonly associated with discovery claims. The SNR values for the mass ranges of interest are also shown in Fig.~\ref{fig:snr}. Given the absence of statistically significant candidates, we report a null result for the current experiment. A Bayesian statistical framework is employed to interpret the data, and upper limits are placed on the coupling constant $g_{\rm s}$ as described below.

Performing a change of variables, we define the normalized quantities
\begin{equation}
    P = \frac{\hat{S}}{\sigma} \quad \text{and} \quad P_{\rm s} = \frac{2 T_{\rm obs} \left| \vec{R}(f_{\rm s}, \mathcal{A}_{\rm s}) \right|^2 g_{\rm s}^2}{\sigma} \;,
\end{equation}
where $P$ represents the normalized excess signal power, and $P_{\rm s}$ corresponds to the normalized signal PSD. Using these definitions, we apply Bayes’ theorem to derive the posterior distribution for the coupling constant $g_{\rm s}$. The resulting expression is given by
\begin{align}
    p(P_{\rm s}^0 \mid P) &\propto p(P_{\rm s}^0) \int p(P_{\rm s} \mid P_{\rm s}^0) \, \mathcal{L}(P \mid P_{\rm s}) \, \mathrm{d}P_{\rm s} 
    \nonumber\\
    &= \frac{e^{-P / (1 + P_{\rm s}^0)}}{(1 + P_{\rm s}^0)^2} \;,
    \label{eq:posterior}
\end{align}
where $\mathcal{L}(P \mid P_{\rm s})$ is the likelihood function for the observed power, given by \cite{1975ApJS...29..285G,Centers:2019dyn},
\begin{equation}
    \mathcal{L}(P \mid P_{\rm s}) = e^{-(P + P_{\rm s})} I_0\!\left(2 \sqrt{P P_{\rm s}}\right) \;,
    \label{eq:likelihood}
\end{equation}
and $I_0$ is the modified Bessel function of the first kind.

Due to the virialized nature of scalar field dark matter, the signal PSD $P_{\rm s}$ is not deterministic. As discussed previously, on timescales much shorter than the coherence time $\tau$, the dark matter field coherently oscillates at the Compton frequency, causing the amplitude $\mathcal{A}_{\rm s}$—and thus the local dark matter density $\rho$—to fluctuate around a mean value $\mathcal{A}_{\rm s}^0 = \sqrt{2 \rho_0} / m_s$, as described in \cite{OHare:2017yze,Centers:2019dyn}. Consequently, we marginalize the likelihood over the distribution of the excess signal PSD, which is modeled as
\begin{equation}
    p(P_{\rm s} \mid P_{\rm s}^0) = \frac{1}{P_{\rm s}^0} e^{-P_{\rm s} / P_{\rm s}^0} \;,
    \label{eq:phi0_prob}
\end{equation}
where $P_{\rm s}^0$ corresponds to the expected signal PSD evaluated at $\mathcal{A}_{\rm s} = \mathcal{A}_{\rm s}^0$, i.e., $P_{\rm s}^0 = P_{\rm s} \big|_{\mathcal{A}_{\rm s} = \mathcal{A}_{\rm s}^0}$.

The final ingredient required for the posterior is the prior distribution for $P_{\rm s}^0$. Following the approach of \cite{Centers:2019dyn,Fedderke:2021rrm,Arza:2021ekq}, we adopt Jeffreys’ prior, given by
\begin{equation}
    p(P_{\rm s}^0) \propto \frac{1}{1 + P_{\rm s}^0} \;.
    \label{eq:prior}
\end{equation}
The posterior distribution $p(P_{\rm s}^0 \mid P)$ is then properly normalized according to
\begin{equation}
    \int_0^\infty p(P_{\rm s}^0 \mid P) \, \mathrm{d}P_{\rm s}^0 = 1 \;.
\end{equation}

For the null search, we derive a 95\% confidence level (CL) upper limit on the coupling constant $g_{\rm s}$ by solving the integral equation
\begin{equation}
    \int_0^{P_{\rm th}} p(P_{\rm s}^0 \mid P) \, \mathrm{d}P_{\rm s}^0 = 0.95 \;,
    \label{eq:upper_limit_integral}
\end{equation}
where $P_{\rm th}$ is the threshold value of $P_{\rm s}^0$ corresponding to the 95\% CL. The coupling constant $g_{\rm s}$ is then obtained from $P_{\rm th}$ via the relation in Eq.~\eqref{eq:psd}. We have verified that, compared to the case where the likelihood in Eq.~\eqref{eq:likelihood} is used directly (i.e., without marginalizing over the stochastic signal amplitude), the inclusion of the stochastic effects through the marginalized likelihood weakens the resulting constraint by approximately a factor of 2.7, consistent with expectations for our short observation time $T_{\rm obs} \ll \tau$, as discussed in \cite{Centers:2019dyn}.

\begin{figure*}[ht]
    \centering
    \includegraphics[width=0.46\linewidth]{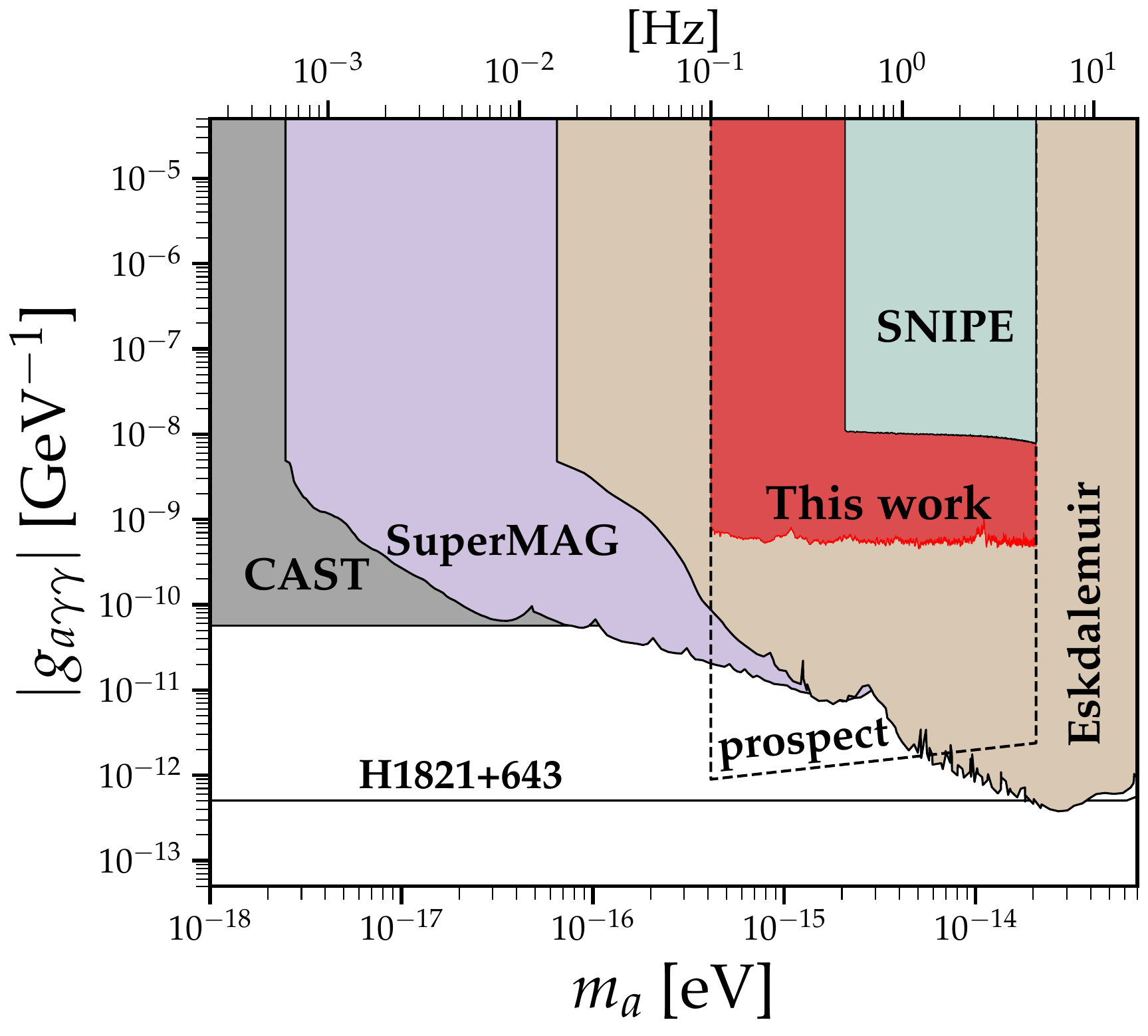}
    \includegraphics[width=0.45\linewidth]{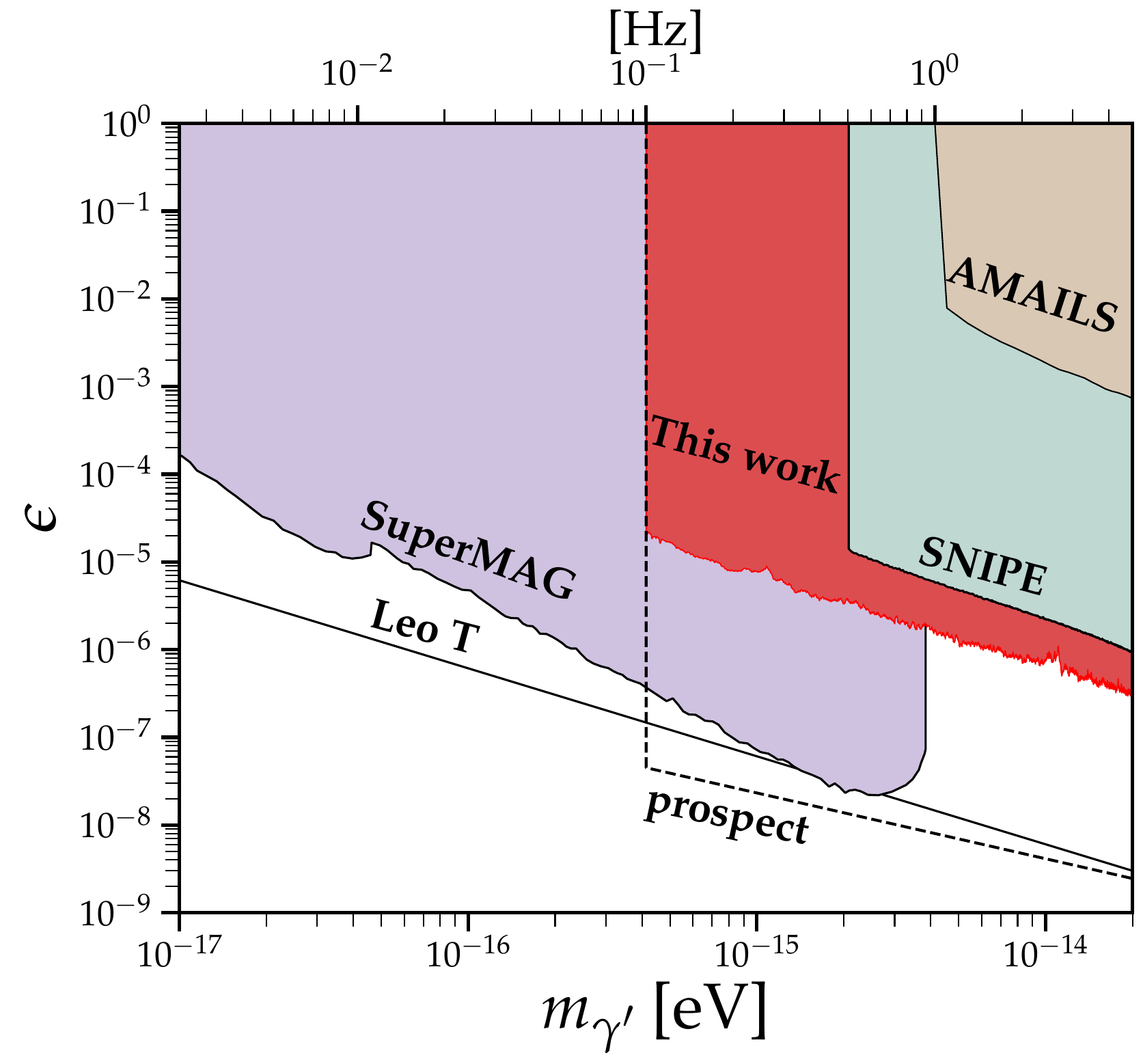}
 \caption{
    \textbf{Left:} 95\% CL limit on $g_{a\gamma\gamma}$ vs $m_a$ from GPEX, compared with SuperMAG~\cite{Arza:2021ekq,Friel:2024shg}, SNIPE-Hunt~\cite{Sulai:2023zqw}, Eskdalemuir~\cite{Taruya:2025zql,Nishizawa:2025xka}, CAST~\cite{CAST:2017uph}, and the quasar H1821+643 bound~\cite{Reynes:2021bpe}. 
    \textbf{Right:} 95\% CL limit on $\epsilon$ vs $m_{\gamma'}$ from GPEX, compared with SuperMAG~\cite{Fedderke:2021aqo,Fedderke:2021rrm,Friel:2024shg}, SNIPE-Hunt~\cite{Sulai:2023zqw}, AMAILS~\cite{Jiang:2023jhl}, and the Leo T dwarf bound~\cite{Wadekar:2019mpc}. Prospective sensitivity for 10 magnetometers ($0.54~\mathrm{pT/\sqrt{Hz}}$, 1 month) is also shown.
}
    \label{fig:limit}
\end{figure*}

Our results, presented in Fig.~\ref{fig:limit}, shows the 95\% CL exclusion limits on the couplings $g_{a\gamma\gamma}$ and $\epsilon$ over the frequency range $0.1 \leq f \leq 5\,\mathrm{Hz}$. This corresponds to an axion or dark photon mass range from $3.5 \times 10^{-16}\, \mathrm{eV}$ to $1.8 \times 10^{-14}\, \mathrm{eV} \;$. Compared to existing constraints from geomagnetic field observations, SuperMAG~\cite{Arza:2021ekq,Fedderke:2021aqo,Fedderke:2021rrm}, SNIPE-Hunt~\cite{Sulai:2023zqw}, Eskdalemuir Observatory~\cite{Nishizawa:2025xka}, our bounds on axions are approximately one order of magnitude stronger than those from SNIPE-Hunt. However, they are less stringent than limits set by CAST \cite{CAST:2017uph}, SuperMAG, and Eskdalemuir, primarily due to their substantially longer observation times. For dark photons, our constraints improve upon those from SNIPE-Hunt and AMAILS~\cite{Jiang:2023jhl}, yet remain weaker than the SuperMAG \cite{Fedderke:2021aqo,Friel:2024shg} results for similar reasons. We note that concurrent with the completion of this manuscript, new experimental constraints on dark photons have been reported by two independent groups: one analyzing archival data from the Eskdalemuir observatory \cite{Nomura:2025rfi}, and another utilizing a dedicated laboratory experiment with a two-sensor array of optically pumped scalar magnetometers \cite{Zhao:2025kdw}.

Additional astrophysical observations provide even more stringent constraints for both axions and dark photons in this parameter space~\cite{Yuan:2020xui,Caputo:2022mah,Wouters:2013hua,Dessert:2020lil,Marsh:2017yvc,Guo:2023hyp,Song:2024rru,Xue:2024zjq,Guo:2025dlk,Fischbach:1994ir,Marocco:2021dku,Bhoonah:2018gjb,Su:2021jvk}. For axions, limits from X-ray observations of the quasar H1821+643 with the Chandra telescope are particularly strong~\cite{Reynes:2021bpe}. For dark photons, the most stringent constraints currently come from analyses of heating and cooling effects induced in the gas-rich Leo T dwarf galaxy~\cite{Wadekar:2019mpc}. It is important to note, however, that these astrophysical bounds depend significantly on the modeling of the relevant systems and underlying mechanisms, and thus entail substantial uncertainties.

{\it \textbf{Conclusion}.} In this work, we present the first geomagnetic observations conducted by the GPEX collaboration to search for axion and dark photon dark matter signals within the Earth's cavity, arising from their interactions with the geomagnetic field. Operating at a remote experimental site isolated from anthropogenic electromagnetic interference and employing a highly sensitive quantum magnetometer, we performed a one-hour observation that reached the intrinsic noise floor of the instrument. No robust signal was observed, the data were analyzed within a Bayesian framework to derive constraints on the axion-photon and dark photon-photon couplings. Remarkably, by using a single atomic magnetometer and only one hour of measurement, we achieved a sensitivity exceeding that of the SNIPE-Hunt experiment.


As the current experimental reach is constrained by magnetometer sensitivity rather than geomagnetic noise, we plan to establish a high-performance network of quantum magnetometers to search for dark matter signals and achieve world leading sensitivity. Such a network can enhance the effective sensitivity and, through correlated measurements, enable confident discrimination of dark matter signals from various noise sources \cite{Pospelov:2012mt,Derevianko:2013oaa,Roberts:2017hla,Afach:2021pfd,Jiang:2023jhl}. For instance, employing 10 quantum magnetometers with sensitivity of 0.54 $\rm{pT/\sqrt{Hz}}$ \cite{Huang:2025akb}, improved by a factor of 40, could enhance the overall sensitivity by about three orders of magnitude after one month of integration, as shown in Fig. \ref{fig:limit}. We anticipate that a future network of quantum magnetometers in remote locations will have the sensitivity to surpass constraints set by SuperMAG, Eskdalemuir, and eventually even astrophysical bounds. Furthermore, as mentioned in \cite{Sulai:2023zqw,Bloch:2023wfz}, employing local multi-sensor arrays to measure the curl of the local magnetic field may extend the accessible frequency range to approximately 1 kHz.

{\it \textbf{Acknowledgement}}. We thank Saarik Kalia and Daniel Gavilan-Martin for the valuable discussions on the data analysis. L.W. is supported by the NNSFC under Grant No. 12275134 and No. 12335005. This work is supported by the National Key Research and Development Program of China (Grant No. 2022YFA1605000), and Quantum Science and Technology-National Science and Technology Major Project (Grant No.~2024ZD0302300).  J.S. is supported by Peking University under startup Grant No. 7101302974 and the National Natural Science Foundation of China (NNSFC) under Grants No. 12025507, No.12450006. J.G. is supported  by the NNSFC under Grant No. 12305111. B.Z. is supported by the NNSFC under Grant No. 12275232 and Shangdong Provincial Natural Science Foundation for Distinguished Young Scholars under Grant No. ZR2025QA20. Q.Y. is supported by the Project for Young Scientists in Basic Research of Chinese Academy of Sciences under Grant No. YSBR-061.

\bibliography{refs}

\onecolumngrid
\clearpage

\setcounter{page}{1}
\setcounter{equation}{0}
\setcounter{figure}{0}
\setcounter{table}{0}
\setcounter{section}{0}
\setcounter{subsection}{0}
\renewcommand{\theequation}{S.\arabic{equation}}
\renewcommand{\thefigure}{S\arabic{figure}}
\renewcommand{\thetable}{S\arabic{table}}
\renewcommand{\thesection}{\Roman{section}}
\renewcommand{\thesubsection}{\Alph{subsection}}

\newcommand{\ssection}[1]{
    \addtocounter{section}{1}
    \section{\thesection.~~~#1}
    \addtocounter{section}{-1}
    \refstepcounter{section}
}
\newcommand{\ssubsection}[1]{
    \addtocounter{subsection}{1}
    \subsection{\thesubsection.~~~#1}
    \addtocounter{subsection}{-1}
    \refstepcounter{subsection}
}
\newcommand{\fakeaffil}[2]{$^{#1}$\textit{#2}\\}

\thispagestyle{empty}
\begin{center}
    \begin{spacing}{1.2}
        \textbf{\large
            \hypertarget{sm}{Supplementary Materials:} Search for Ultralight Dark Matter with Quantum Magnetometry in the Earth’s Cavity\\
        }
    \end{spacing}
    \par\smallskip
    Ariel Arza,$^{1,2}$
    Yuanlin Gong,$^{1}$
    Jun Guo,$^{3}$
    Xiaofei Huang,$^{4}$
    Jing Shu,$^{5,6,7}$
    Hongliang Tian,$^{1}$
    Wenyu Wang,$^{8}$
    Kai Wei,$^{4,9}$
    Lei Wu,$^{1,2}$
    Mingming Xia,$^{10}$
    Jin Min Yang,$^{11,12}$
    Qiang Yuan,$^{13,14}$
    Yang Zhang$^{11}$
    Yi Zhang,$^{13,14}$
    Bin Zhu,$^{15}$
    \par
    {\small
        \fakeaffil{1}{Department of Physics and Institute of Theoretical Physics, Nanjing Normal University, Nanjing, 210023, China}
        \fakeaffil{2}{Nanjing Key Laboratory of Particle Physics and Astrophysics, Nanjing, 210023, China}
        \fakeaffil{3}{College of Physics and Communication Electronics, Jiangxi Normal University, Nanchang 330022, China}
        \fakeaffil{4}{School of Instrumentation and Optoelectronic Engineering, Beihang University, Beijing 100191, China}
        \fakeaffil{5}{School of Physics and State Key Laboratory of Nuclear Physics and Technology, Peking University, Beijing 100871, China}
        \fakeaffil{6}{Center for High Energy Physics, Peking University, Beijing 100871, China}
        \fakeaffil{7}{Beijing Laser Acceleration Innovation Center, Huairou, Beijing, 101400, China}
        \fakeaffil{8}{Beijing University of Technology, 100124, Beijing, China}
        \fakeaffil{9}{Quantum Science and Technology College, Beihang University, Beijing 100191, China}
        \fakeaffil{10}{Hangzhou Institute of Extremely-Weak Magnetic Field Major National Science and Technology Infrastructure, Hangzhou, 310051, China}
        \fakeaffil{11}{Center for Theoretical Physics, Henan Normal University, Xinxiang 453007, China}
        \fakeaffil{12}{Institute of Theoretical Physics, Chinese Academy of Sciences, Beijing 100190, China}  
        \fakeaffil{13}{Key Laboratory of Dark Matter and Space Astronomy, Purple Mountain Observatory,
        Chinese Academy of Sciences, Nanjing 210008, China}
        \fakeaffil{14}{School of Astronomy and Space Science, University of Science and Technology of China,
        Hefei 230026, China}
        \fakeaffil{15}{School of Physics, Yantai University, Yantai 264005, China}
    }

\end{center}
\par\smallskip

\section{Distribution of the Noise Amplitude}

\begin{figure*}[ht]
    \centering
    \includegraphics[width=0.45\linewidth]{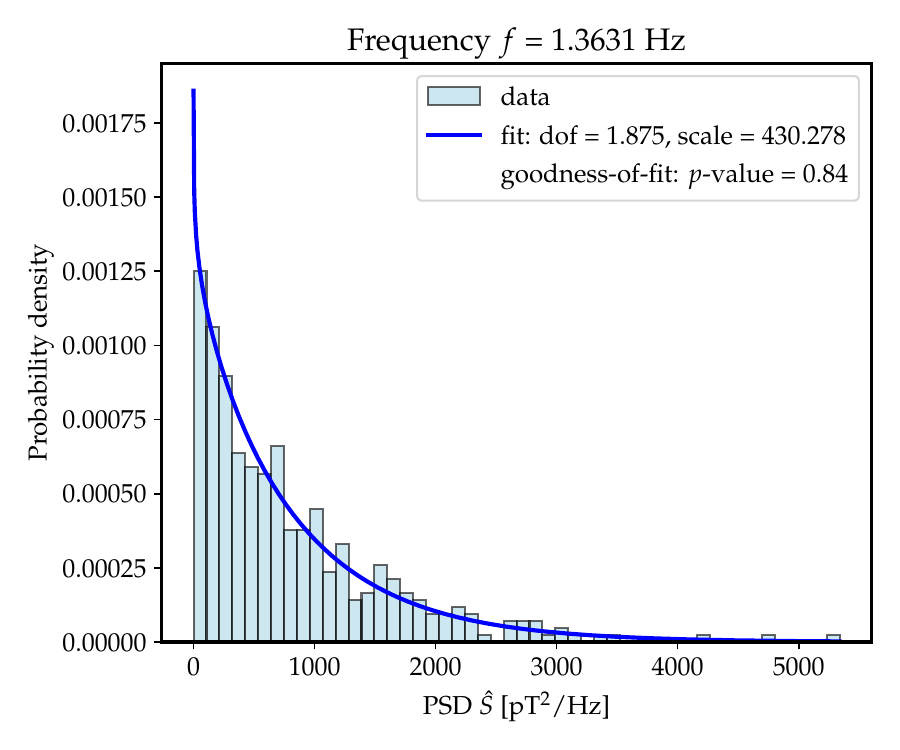}
    \includegraphics[width=0.45\linewidth]{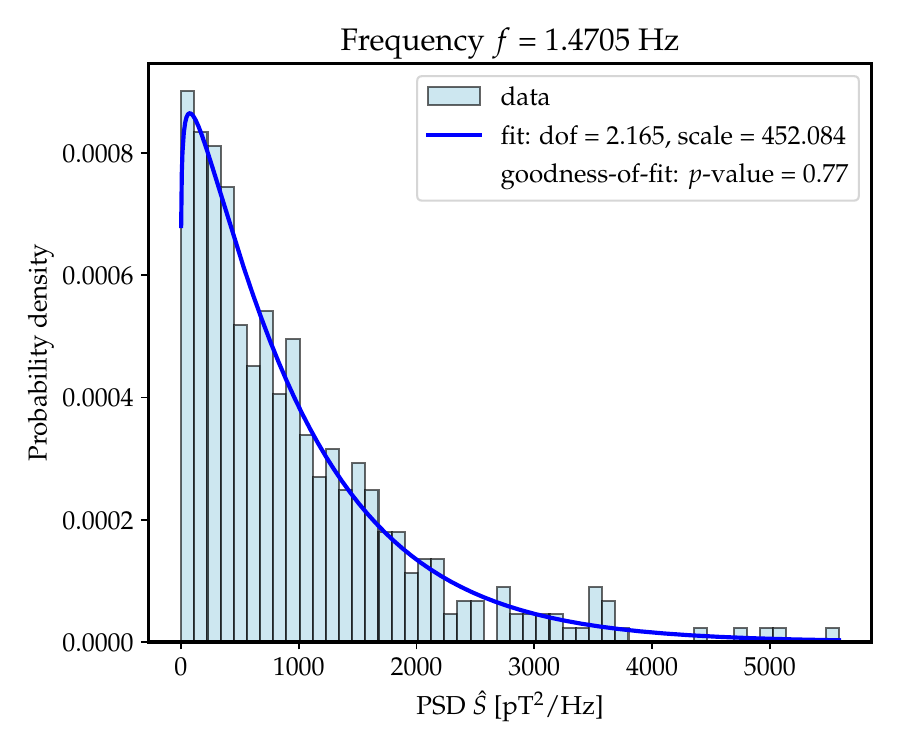}\\
    \includegraphics[width=0.45\linewidth]{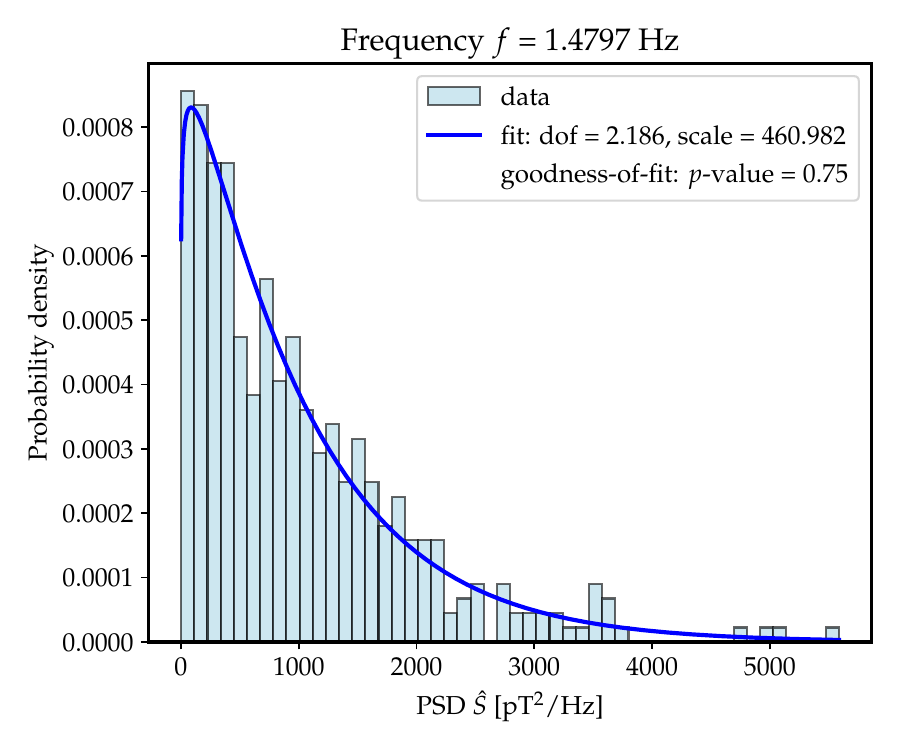}
    \includegraphics[width=0.45\linewidth]{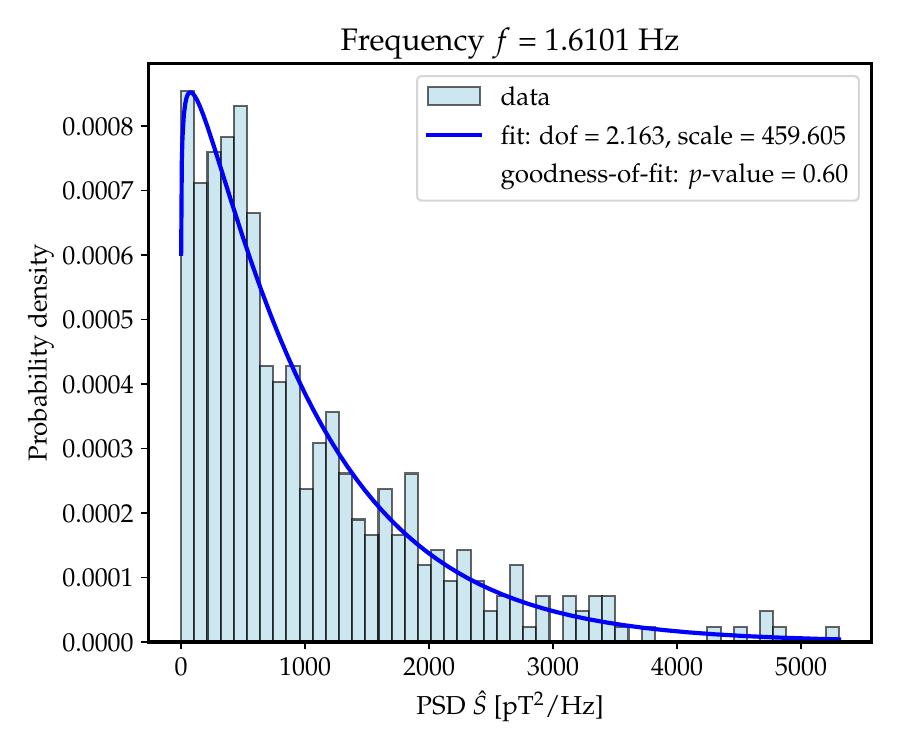}\\
    \includegraphics[width=0.45\linewidth]{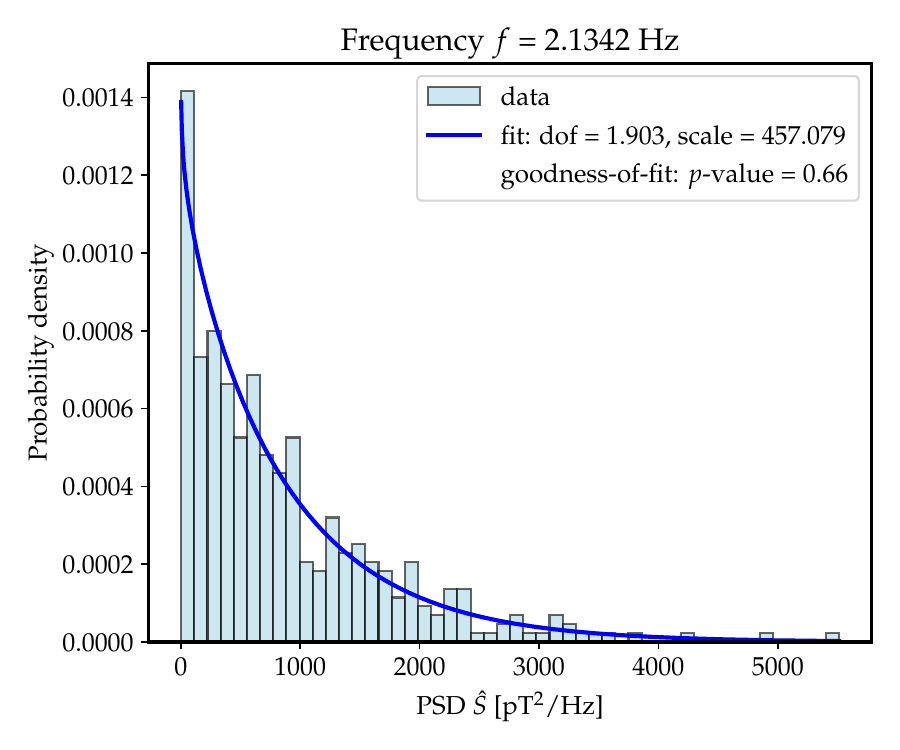}
    \includegraphics[width=0.45\linewidth]{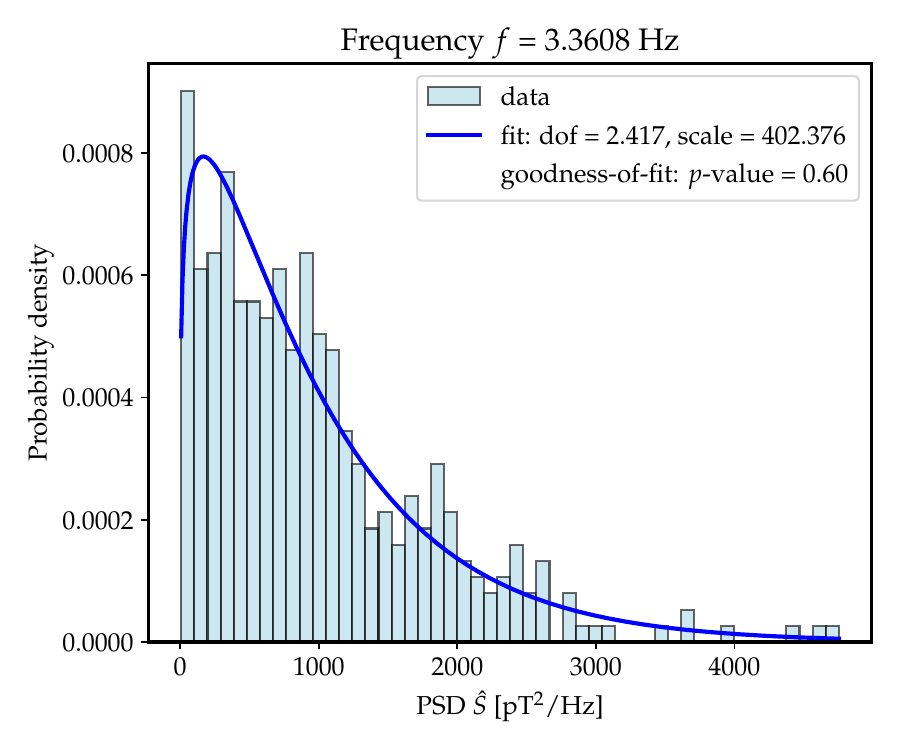}    
    \caption{Six representative noise amplitude distributions from observed PSD at specific frequencies. The blue curves denote the best-fit $\chi^2$ distributions, along with the corresponding fitted dofs, scale parameters, and $p$-values from the goodness-of-fit test.}
    \label{fig:pdf}
\end{figure*}

\begin{figure*}[ht]
    \centering
    \includegraphics[width=0.5\linewidth]{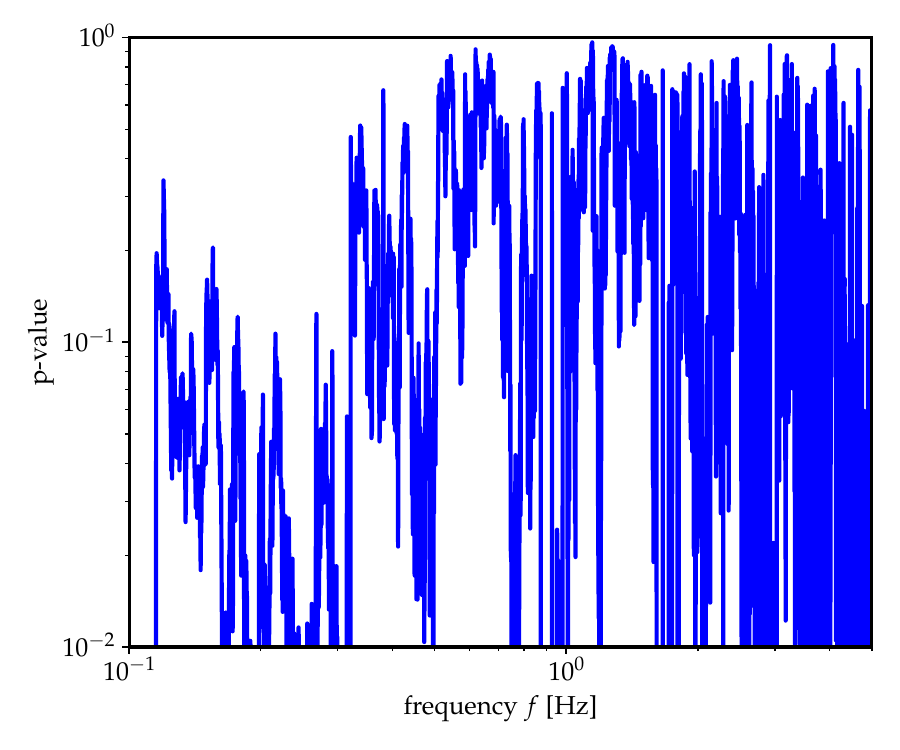}
    \caption{$p$-value from the $\chi^2$ goodness-of-fit test for the frequency range of interest, 0.1 Hz to 5 Hz.}
    \label{fig:goodness}
\end{figure*}

As discussed in the main text, the noise power spectral density (PSD) follows a $\chi^2$ distribution with two degrees of freedom (dof). The flat noise behavior across our frequency range of interest allows us to study the distribution of noise amplitudes and estimate the noise variance using the PSD values of neighboring frequency bins. In Fig.~\ref{fig:pdf}, we show the distributions of the observed PSD within $\pm 200$ frequency bins around $f = f_{\rm s}$, excluding the three bins centered on $f_{\rm s}$, where $f_{\rm s}$ denotes the expected signal frequency. For the six representative frequencies---1.3631, 1.4705, 1.4797, 1.6101, 2.1342, and 3.3608~Hz---the distributions are well described by a $\chi^2$ distribution with a dof close to two, as shown in each subplot. We also present the $p$-values from the $\chi^2$ goodness-of-fit test~\cite{ParticleDataGroup:2018ovx} and the corresponding scale parameters in the same figure. Furthermore, Fig.~\ref{fig:goodness} shows the $p$-values of the goodness-of-fit test across the entire frequency range of interest.

\end{document}